\documentclass[namedreferences]{solarphysics}
\usepackage[optionalrh,solaromanenum]{spr-sola-addons} 
\usepackage{graphicx}                    
\usepackage{color}                       
\usepackage{url}                         

\usepackage[pdfborder={0 0 0},colorlinks=true,urlcolor=blue,breaklinks]
{hyperref}
\newcommand{\doiurl}[1]{\href{http://dx.doi.org/#1}{#1}}
\newcommand{\adsurl}[1]{\href{http://adsabs.harvard.edu/abs/#1}{#1}}

\def\be{\begin{equation}}
\def\ee{\end{equation}}
\def\bea{\begin{eqnarray}}
\def\eea{\end{eqnarray}}

\def\curl{{\bf \nabla}\times}
\def\grad{{\bf \nabla}}
\def\div{{\bf \nabla}\cdot}
\def\half{{\textstyle{1\over2}}}

\newfont{\myfont}{cmmib10}

\newcommand{\bomega}{\hbox{\myfont \symbol{33} }}

\def\bi{\bf}
\def\cf{{\it cf}.}
\def\etc{{\it etc}.}

\newcommand{\aap}{    {\it Astron. Astrophys.}}

\newcommand{\apj}{    {\it Astrophys. J.}}

\newcommand{\grl}{    {\it Geophys. Res. Lett.}}

\newcommand{\jgr}{    {\it J. Geophys. Res.}}

\newcommand{\pasj}{   {\it Pub. Astron. Soc. Japan}}

\newcommand{\solphys}{{\it Solar Phys.}}
 
\newcommand{\ssr}{    {\it Space Sci. Rev.}}

\begin{document}

\begin{article}

\begin{opening}

\title{Transfer of Energy, Potential, and Current by Alfv\'en Waves in Solar Flares}

%
\author{D.B.~Melrose{}$^{1}$\sep  M.S.~Wheatland{}$^{1}$
       }

%
\runningauthor{D.B. Melrose, M.S. Wheatland}
\runningtitle{Alfv\'en Waves in Solar Flares}

%
  \institute{$^{1}$ Sydney Institute for Astronomy, School of Physics, University of Sydney, NSW 2006, Australia
email: \href{mailto:melrose@physics.usyd.edu.au}{melrose@physics.usyd.edu.au}, \href{mailto:m.wheatland@physics.usyd.edu.au}{m.wheatland@physics.usyd.edu.au}
             }

\begin{abstract}
Alfv\'en waves play three related roles in the impulsive phase of a solar flare:  they transport energy from a generator region to an acceleration region; they map the cross-field potential (associated with the driven energy release) from the generator region onto the acceleration region; and within the acceleration region they damp by setting up a parallel electric field that accelerates electrons and transfers the wave energy to them. The Alfv\'en waves may also be regarded as setting up new closed current loops, with field-aligned currents that close across field lines at boundaries. A model is developed for large-amplitude Alfv\'en waves that shows how Alfv\'en waves play these roles in solar flares. A picket-fence structure for the current flow is incorporated into the model to account for the ``number problem'' and the energy of the accelerated electrons.
\end{abstract}

%
\keywords{solar flares; Alfv\'en waves; electron acceleration}

\end{opening}

%
\section{Introduction}
\label{s:introduction} 

There are long-standing, unsolved problems in the physics of solar flares. A large fraction of the magnetic energy released in a flare appears in $\varepsilon=10\,-\,20\,$keV electrons that produce hard X-rays and type~III solar radio bursts. One problem is that there is no satisfactory model for the acceleration of these electrons, which,  in the older literature, was referred to as ``first phase'' acceleration \cite{WSW} and as ``bulk energization'' of electrons. Runaway acceleration \cite{H85} due to a parallel electric field [$E_\parallel$] seems the only viable  acceleration mechanism, but there is no accepted model for how $E_\parallel\ne0$ is set up and maintained in an acceleration region. There is also a long-standing ``number problem'' \cite{Hetal76,MB89,BKB10} that can be expressed in various ways; for example, the rate, $\,{\dot{\!N}}>10^{36}\rm\,s^{-1}$, of precipitation of accelerated electrons inferred from hard X-ray observations implies a total of $10^{39}$ accelerated electrons in a flare of duration $10^3\,$seconds, and this exceeds the number of electrons stored in the entire flaring flux loop, estimated to be $10^{37}$ \cite{EH95}. Another problem is that the power released in a magnetic explosion \cite{M12a} may be expressed as the product [$I\Phi$] of a current and a potential, and if one assumes a current of order $I=10^{11}\,$A, consistent with vector magnetogram data, then to account for the energy released one requires a very large potential, $\Phi=10^{10}\,$V, for which there is no direct evidence. The latter two problems are related, in that there is a single large unexplained factor [$M$] with $e\,{\dot{\!N}}=MI$ and $e\Phi=M\varepsilon$, suggesting $M$ of order $10^6$. To account for the factor $M$,  \inlinecite{H85} argued that there are multiple current paths, with the current [$I$] flowing up and down $M$ times. However, no mechanism has been proposed to account for this seemingly bizarre current pattern. The resupply of electrons is attributed to a return current \cite{BB84,SS84,vdO90,LS91,EH95}. Arguments related to the number problem and the return current suggest an acceleration region near (or in) the chromosphere, where there is an adequate supply of electrons.

\begin{figure} [t]
\centerline{
\includegraphics[scale=0.215]{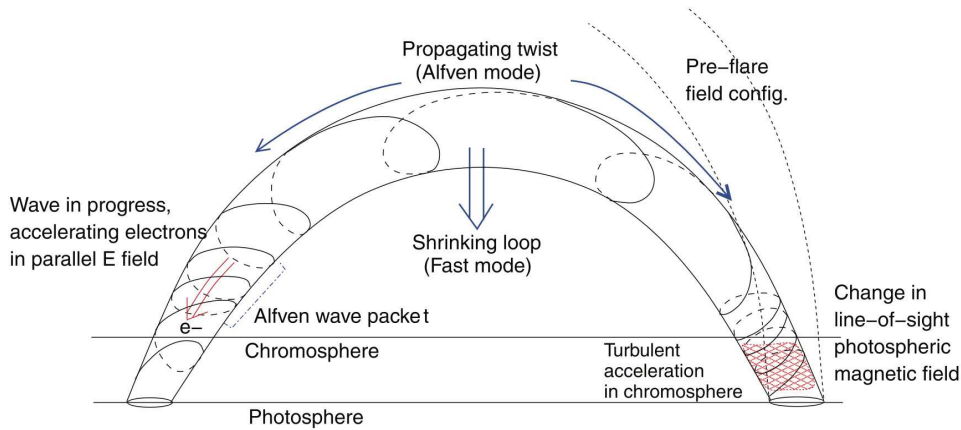}
}
\caption{A cartoon showing a model due to Fletcher and Hudson (2008) for energy release in a flare in which energy is transported from an energy-release site (effectively at the top of the loop), via a propagating twist to an acceleration region in the chromosphere. In the Alfv\'en model in this article, the flux loop is replaced by a vertical magnetic field, with the energy-release site at the top.
}
\label{fig:FH08}
\end{figure} 

A model developed by \inlinecite{FH08} addresses the number problem. The model is shown in Figure~\ref{fig:FH08}. Energy release in the corona due to magnetic reconnection generates large-scale Alfv\'en wave pulses, which propagate within a loop to the chromosphere where they damp and accelerate electrons via a turbulent cascade. The model includes some of the essential features in an acceptable flare model. Notably the energy-release site and the acceleration/dissipation region are remote from each other, with transport of energy between them via Alfv\'en waves.

The model envisaged here \cite{M12b} for energy release involves a generator (or energy-release) region, where reconnection allows conversion of inflowing magnetic energy into outflow in the form of plasma kinetic energy and an Alfv\'en flux (\cf\ \opencite{FH08}). We do not discuss the details of the generator region here, but several comments are appropriate. The energy release needs to be driven, and one possible mechanical driver is a pressure gradient, associated with collection of dense plasma near the top of the flaring loop (\citeauthor{H01},  \citeyear{H01}, \citeyear{H09};  \citeyear{H12}, \opencite{FH08}). In this case one needs to consider the energy and momentum associated with this dense plasma, and how it is transferred to Alfv\'en waves. Our view (\citeauthor{M12a}, \citeyear{M12a}, \citeyear{M12b}) is that the driver is the Maxwell stress, and that this becomes available as reconnection allows the magnetic figuration to change to a less stressed state. The energy and momentum in the plasma then play only a minor role. We identify the power [$I\Phi$] released to redirection of pre-existing current across field lines in the generator region, with the electromotive force [$\Phi$] attributed to the rate of change of the stored magnetic flux. This cross-field current [$I$] is assumed to be redirected so that it flows along field lines through an acceleration region, and closes across field lines in the photosphere, and then flows back to the generator region. 

\begin{figure} [t]
\centerline{
\includegraphics[scale=0.3]{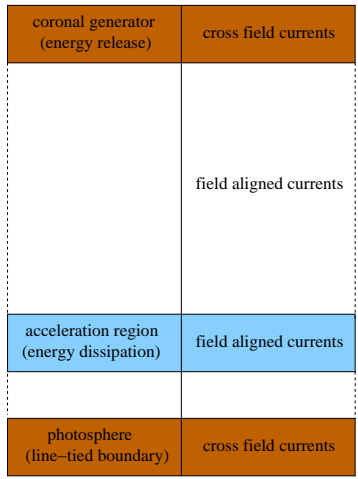}
}
\caption{A schematic diagram of the model adopted in this article, as explained in the text.
}
\label{fig:schematic}
\end{figure}

A schematic diagram of the new model is shown in Figure~\ref{fig:schematic}. The magnetic field is assumed to be vertical, with the top of the figure in the corona, and the bottom at the photosphere. The redirection of current at the coronal generator is analogous to the redirection of current that occurs in the so-called current wedge in a  magnetospheric substorm \cite{McP73,PHT02}. The physics underlying this effect is related to a conducting boundary \cite{S55}, which is the plasma wall in a laboratory context, the ionosphere in the magnetospheric context, and the photosphere in the context of a flare. Specifically, when a current is driven across field lines in the body of the plasma, it closes by flowing along the field lines to a region where the conductivity allows it to close across field lines. Following  \inlinecite{H85}, it is assumed that this redirection occurs $M$ times in a flare, so that the potential imposed at the generator region is $\Phi/M$ for each such redirected current path.

In this article an analytic model for the large-scale Alfv\'en waves is developed and used to describe various roles that Alfv\'en waves play in such a model.  The features that are new, in the solar context, are the transport of potential along field lines, and the setting up of new closed current loops. The electric field in the wave is expressed in terms of a wave potential, which also determines the field-aligned current (FAC) density in the wave. The waves map the potential, imposed at the generator region, onto the acceleration region, where the waves are assumed to damp, converting the cross-field potential at the top of the acceleration region into a field-aligned potential within the acceleration region.  Figure~\ref{fig:pot} illustrates the mapping of the potential and the role of the acceleration region. With the magnetic field in the vertical direction, the equipotentials shown by dashed lines are vertical above the acceleration region. The potential set up in the generator region (above the top of the diagram) maps along field lines to the acceleration region, where a field-aligned potential is established. The field-aligned electric field [$E_\parallel$] in the region extracts energy from the waves and transfers it to electrons. 

Assuming that current is conserved in an Alfv\'en wave ($\div{\bi J}=0$), an Alfv\'en wave can transfer a FAC from one field line to another, and can transfer a cross-field current from one location to another, but it cannot introduce any intrinsically new current. This implies that an Alfv\'en wave that is localized to a flux tube can have no net FAC: the FAC in the wave must change sign across the wavefront, and include equal up and down currents. The up and down FACs in the wave close across field lines in the generator region and in the photosphere, forming a closed current loop, and transferring the cross field current from one region to the other.

The general model for Alfv\'en waves is developed in Section~\ref{s:Alfven}, and some relevant  boundary conditions are discussed in Section~\ref{s:boundary}. The model is applied to cylindrical and planar geometries in Section~\ref{s:cylindrical}. Some aspects of the extension of the model into the acceleration region are discussed in Section~\ref{s:dissipation}, and a general discussion of the results is presented in Section~\ref{s:discussion}. The conclusions are summarized in Section~\ref{s:conclusion}.

\section{Alfv\'en Waves}
\label{s:Alfven} 

The model for Alfv\'en wave adopted here is based on one developed in connection with magnetospheric substorms \cite{VH98,PHT02}. Conventional treatments of Alfv\'en waves are based on MHD or the low-frequency limit of cold-plasma theory, and assume that the waves are of small amplitude. The approach adopted here applies to Alfv\'en waves even when neither of these theories is valid, and there is no restriction on the wave amplitude. Our approach is based on two well-known results from orbit theory: the electric and polarization drifts \cite{N63}. It allows one to include the plasma response to an electric field (and its time derivative) without assuming that its amplitude is small.

\subsection{Electric and Polarization Drifts}

Any electric field, including both the inductive electric field due to the changing magnetic field, and the electric field in an Alfv\'en wave, can be separated into components perpendicular [${\bi E}_\perp$] and parallel [$E_\parallel{\hat{\bi z}}$] to the background magnetic field [${\bi B}=B{\hat{\bi z}}$] which is assumed uniform along the $z$~axis. Subject to relatively weak conditions (on the slowly time-varying electric field), one has 
\be
{\bi E}={\bi E}_\perp+E_\parallel{\hat{\bi z}},
\qquad
{\bi E}_\perp=-{\bi u}\times{\bi B},
\label{pd1}
\ee 
with  ${\bi u}={\bi E}_\perp\times{\bi B}/B^2$ the electric drift velocity. In MHD, ${\bi u}$ is interpreted as the fluid velocity, and Equation (\ref{pd1}) is attributed to Ohm's law for infinite conductivity, which also implies $E_\parallel=0$. The interpretation of ${\bi u}$ as the electric drift velocity is different from the usual MHD interpretation: here ${\bi E}$ implies ${\bi u}$ and $E_\parallel$ is arbitrary, whereas in MHD ${\bi u}$ implies ${\bi E}$ with $E_\parallel=0$.  The divergence of the electric field, in the form Equation (\ref{pd1}), gives the charge density,
\be
{\rho\over\varepsilon_0}=-B{\hat{\bi z}}\cdot\bomega+{\partial E_\parallel\over\partial z},
\qquad
\bomega=\curl{\bi u},
\label{pd2}
\ee 
where ${\bomega}$ is the vorticity in a fluid interpretation.

A temporally changing (perpendicular) electric field corresponds to a displacement current $\varepsilon_0\partial{\bi E}_\perp/\partial t$, and this causes a polarization drift, which is along $\partial{\bi E}_\perp/\partial t$ at a velocity proportional to the mass to charge ratio. After summing over the contributions of all species of charged particles, this implies a polarization current density
\be
{\bi J}_{\rm pol \perp}={c^2\over v_{\rm A}^2}{\bi J}_{\rm displ},
\qquad
{\bi J}_{\rm displ}=\varepsilon_0{\partial{\bi E}_\perp\over\partial t}.
\label{pd3}
\ee
The sum of the two currents in Equation (\ref{pd3}) is $c^2/v_0^2$ times ${\bi J}_{\rm displ}$, where $v_0=v_{\rm A}/(1+v_{\rm A}^2/c^2)^{1/2}$ is the MHD speed and $v_{\rm A}$ is the conventional Alfv\'en speed. 

Equation (\ref{pd3}) may also be derived using cold-plasma theory, as shown in Appendix~\ref{A-appendix}.

\subsection{Wave Equation}

Including the polarization current in Amp\`ere's equation, and combining it with Faraday's equation, leads to a wave equation for the electric field. The parallel component of the cold-plasma response is given by Equation (\ref{pd8}), and including it in the parallel component of the wave equation gives
\be
\nabla^2{\bi E}_\perp+{\hat{\bi z}}\nabla^2E_\parallel-{\grad\rho\over\varepsilon_0}
={1\over v_0^2}{\partial^2{\bi E}_\perp\over\partial t^2}
+{\hat{\bi z}}\left[{\omega_p^2\over c^2}E_\parallel
+{1\over c^2}{\partial^2E_\parallel\over\partial t^2}\right],
\label{pd9a}
\ee
\be
{\grad\rho\over\varepsilon_0}
=\left({\bf\nabla}_\perp
+{\hat{\bi z}}{\partial\over\partial z}\right)
\left[{\bf\nabla}_\perp\cdot{\bi E}_\perp+{\partial E_\parallel\over\partial z}\right].
\label{pd9b}
\ee
For $E_\parallel\ne0$ the term involving the charge density couples the perpendicular and parallel components in Equation (\ref{pd9a}). 

The wave equation Equation (\ref{pd9a}) simplifies for $E_\parallel=0$ and ${\bf\nabla}_\perp\times{\bi E}_\perp=0$, with the latter condition implying ${\bf\nabla}_\perp({\bf\nabla}_\perp\cdot{\bi E}_\perp)=\nabla^2_\perp{\bi E}_\perp$. Then Equation (\ref{pd9a}) becomes the usual wave equation
\be
\left[{1\over v_0^2}{\partial^2\over\partial t^2}-{\partial^2\over\partial z^2}\right]{\bi E}_\perp
=0.
\label{MEperpt}
\ee

\subsection{Wave Potential}

A general solution of  Equation (\ref{MEperpt}) is of the form
\be
{\bi E}_\perp(x,y,z,t)={\bi E}^+_\perp(x,y,z+v_0t)+{\bi E}^-_\perp(x,y,z-v_0t).
\label{AW1}
\ee
Assuming the $z$ axis is vertical, the $+$~solution is propagating down, and is referred to as a direct wave, and the $-$~solution is propagating up, and is referred to as a reflected wave. The condition ${\bf\nabla}_\perp\times{\bi E}^\pm_\perp=0$ must be satisfied for Equation (\ref{AW1}) to be valid, and this implies that ${\bi E}^\pm_\perp$ may be written
\be
{\bi E}^\pm_\perp=-{\bf\nabla}_\perp\Phi^\pm.
\label{AW1a}
\ee
The quantity  $\Phi_\perp^\pm$ is interpreted as the wave potential. The two solutions satisfy advection equations:
\be
{\partial\over\partial t}{\bi E}^\pm_\perp
=\pm v_0{\partial\over\partial z}{\bi E}^\pm_\perp.
\label{AW2}
\ee
Using Equation (\ref{AW2}), the magnetic and electric fields in the wave are related by
\be
v_0{\bi B}^\pm_\perp
=\pm {\hat{\bi z}}\times{\bi E}^\pm_\perp.
\label{AW2a}
\ee
The electric drift velocity may be interpreted as the fluid velocity [${\bi u}^\pm={\bi E}^\pm_\perp\times{\bi B}/B^2$] in the wave. It satisfies the Wal\'en relation in the form
\be
{{\bi u}^\pm\over v_0}=\mp{{\bi B}^\pm_\perp\over B},
\label{AW2b}
\ee
where Equation (\ref{AW2a}) is used. 

The energy density in the waves satisfies the well-know equipartition relation for Alfv\'en waves,
\be
\half\varepsilon_0|{\bi E}^\pm_\perp|^2
+\half\eta|{\bi u}^\pm|^2={|{\bi B}^\pm_\perp|^2\over 2\mu_0},
\label{AW2c}
\ee
where $\eta$ is the mass density. 

\subsection{Parallel Current}

It is convenient \cite{SL06} to define an effective current density [${\bi J}'$] that is the sum of the actual current density and the displacement current, 
\be
{\bi J}'={\bi J}+\varepsilon_0{\partial{\bi E}\over\partial t},
\qquad
\div{\bi J}'=0,
\label{AW2d}
\ee
where the latter follows from Maxwell's equations. The perpendicular current associated with the waves is ${\bi J}'_\perp={\bi J}'^+_\perp+{\bi J}'^-_\perp$ with
\be
{\bi J}'^\pm_\perp={1\over\mu_0v_0^2}{\partial{\bi E}^\pm_\perp\over\partial t}
=\pm{1\over R_{\rm A}}{\partial{\bi E}^\pm_\perp\over\partial z},
\label{AW3}
\ee
where Equation (\ref{AW2}) is used, and where $R_{\rm A}=\mu_0v_0$ is the Alfv\'en impedance.  One has
\be
{\partial J'^\pm_\parallel\over\partial z}=-{\bf\nabla}_\perp\cdot{\bi J}'^\pm_\perp
=\mp{1\over R_{\rm A}}{\partial{\bf\nabla}_\perp\cdot{\bi E}^\pm_\perp\over\partial z},
\label{AW4}
\ee
which implies
\be
J'^\pm_\parallel=\mp{1\over R_{\rm A}}{\bf\nabla}_\perp\cdot{\bi E}^\pm_\perp
=\pm{1\over R_{\rm A}}\nabla^2_\perp\Phi_\perp^\pm.
\label{AW5}
\ee
Assuming $E_\parallel=0$, Equation (\ref{AW5}) determines $J_\parallel=J'_\parallel$.

\section{Boundary Conditions}
\label{s:boundary} 

The Alfv\'en wave model shows how energy is transported, potential is transferred and new current loops are set up. In applying this to an idealized flare model, one needs to impose boundary conditions in three regions: the generator region, the acceleration region, and the photosphere. Prior to a flare, the acceleration region does not exist, and its turning on is identified with the triggering of the energy release in a flare. 

\subsection{Reflection Coefficients}

An Alfv\'en wave incident on a boundary between between two plasmas can be partly reflected and partly transmitted. The ``waves'' here have effectively infinite wavelength, in the sense that the boundary regions may be regarded as of zero thickness compared with the wavelength.

In the magnetospheric application, the lower boundary is the ionosphere, which is partially ionized, implying nonzero Pedersen and Hall conductivities. Let the reflection coefficient be $r_{\rm ref}$. The relation between the reflected and incident waves is \cite{S70}
\be
{\bi E}^-_\perp=r_{\rm ref}\,{\bi E}^+_\perp,
\qquad
J^-_\parallel=-r_{\rm ref}\,J^+_\parallel,
\qquad
r_{\rm ref}=-{R_{\rm A}-R_P\over R_{\rm A}+R_P},
\label{AW9}
\ee
where $1/R_P$ is the height-integrated Pedersen conductivity. The reflection coefficient is zero in the case of perfect absorption, and this corresponds to the impedance matching condition $R_P=R_{\rm A}$.

In the solar context, the chromosphere and photosphere are partially ionized, and one could model the effect of the photosphere using Equation (\ref{AW9}) with $R_{\rm A}\gg R_P$. A more relevant effect is the increase in density from above to below the photosphere/chromosphere, implying a decrease in Alfv\'en speed. For long wavelength waves, the reflection coefficient at a boundary between two non-dissipative plasmas, labeled~1 and~2, is
\be
r_{\rm ref}
=-{R_{A1}-R_{A2}\over R_{A1}+R_{A2}},
\label{refl1}
\ee
with $R_{A1,2}=\mu_0v_{A1,2}$. 

The simplest approximation to the photospheric boundary condition corresponds to line-tying, which is the limit of infinite inertia and zero Alfv\'en speed, $R_{A2}\to0$ in Equation (\ref{refl1}). Thus the line-tying boundary condition corresponds to $r_{\rm ref}\to-1$. Line-tying implies ${\bi E}^-_\perp=-{\bi E}^+_\perp$, ${\bi u}^-=-{\bi u}^+$, ${\bi B}^-_\perp={\bi B}^+_\perp$, $J^-_\perp=J^+_\perp$, $J^-_\parallel=J^+_\parallel$. Hence there is no electric field or fluid velocity at the boundary, but there is a net current and associated magnetic field. A stress imposed in the generator region is transferred to the photosphere by the waves, where the resulting ${\bi J}\times{\bi B}$ force is balanced by the assumed infinite inertia. This stress is transferred back to the generator region by the waves.

One also needs a boundary condition at the generator region. Suppose this is $r_{\rm ref}=-1$, which corresponds to a constant-voltage generator \cite{PHT02}. At such a boundary, the electric fields in the direct and reflected waves cancel [${\bi E}_\perp^++{\bi E}_\perp^-=0$] and the parallel currents add [$J_\parallel^+=J_\parallel^-$] with this net current modifying the cross-field current in the generator region. The opposite assumption [$r_{\rm ref}=1$] corresponds to a constant-current generator. This boundary condition corresponds to ${\bi E}_\perp^+=-{\bi E}_\perp^-$ and $J_\parallel^++J_\parallel^-=0$, so that the cross-field current in the generator is unchanged, and the electric field and flow velocity are modified. Intermediate cases between these two limits have been discussed in the literature on substorms \cite{L85,VHG99}. 

Prior to a flare, the stresses are in balance, field lines are equipotentials and there is no energy transport. Any stress imposed on the generator region is prevented from causing the plasma to move (dynamically) by line-tying in the photosphere. 

\subsection{Boundary Condition on the Flux Tube}

Alfv\'en waves transfer current across field lines, allowing redistribution of field aligned current (FAC), but they cannot generate intrinsically new FACs. The redistribution of FAC may be interpreted as a current loop associated with the Alfv\'en waves, with up and down FACs closing across field lines in both the generator region and the conducting boundary.  The net FAC current associated with the Alfv\'en waves, found by integrating $J_\parallel^++J_\parallel^-$ over the cross-sectional area of the flux tube within which the waves are confined, must be zero. Using Equation (\ref{AW1a}) and Equation (\ref{AW5}), this leads to a restriction on the form of $\Phi^++\Phi^-$. Assuming that this restriction applies separately to $\Phi^\pm$, the condition is
\be
\left.{\bi n}\cdot{\bf\nabla}\Phi^\pm\right|_S=0,
\label{BC1}
\ee
where $S$ is the surface of the flux tube and ${\bi n}$ is the normal to it. For example, in a cylindrical model for a flux tube of radius $R$, Equation (\ref{BC1}) requires that ${\mathrm d}\Phi^\pm/{\mathrm d}r=0$ at $r=R$.

\section{Cylindrical and Planar Models}
\label{s:cylindrical} 

Cylindrical and planar models for the system of Alfv\'en waves are developed in this section. For simplicity, impedance matching is assumed, so that there is no reflected wave.

\subsection{Cylindrical Model for the Potential}

A polynomial model for the potential that satisfies Equation (\ref{BC1}) is
\be
{{\mathrm d}\Phi^+_\perp\over {\mathrm d}r}=A r^a(R-r)^b,
\label{RC5}
\ee
with $A,a,b$ constants. It follows from Equation (\ref{AW5}) with Equation (\ref{RC5}) that one has
\be
J_\parallel={1\over R_{\rm A}}A r^{a-1}(R-r)^{b-1}[(a+1)R-(a+b+1)r].
\label{RC6}
\ee
To avoid singularities at $r=0$ and $r=R$ one needs $a\ge1$ and $b\ge1$, respectively. The sign of the FAC reverses at $r=r_0=(a+1)R/(a+b+1)$, such that the direct and return currents flow at $r<r_0$ and $r>r_0$. This dividing radius [$r_0$] decreases with increasing $b$. One may integrate Equation (\ref{RC5}) to find an explicit expression for $\Phi^+(r)$. For $a\ge1$ and $b\ge1$, the potential difference $\Phi^+(r)-\Phi^+(0)$ does not change sign in the region $0<r\le R$.

The Poynting flux at a given $r$ is proportional to $({\mathrm d}\Phi^+/{\mathrm d}r)^2$, which goes to zero for $r\to R$. The total energy flux is determined by integrating the Poynting vector over the flux tube. The direct and reverse currents are found by integrating $J_\parallel$ over the ranges $0<r<r_0$ and $r_0<r<R$, and are equal in magnitude. Both the total energy flux and the direct and return currents increase with increasing $b$, corresponding to decreasing $r_0$.

\subsection{Planar Models for the Potential}

Let the spatial variation perpendicular to the magnetic field be along the $x$ axis, with no dependence on $y$. It is convenient to assume that the Alfv\'en wave propagates between boundaries at $x=x_1$ and $x=x_2$.

A polynomial model that satisfies Equation (\ref{BC1}) is
\be
{{\mathrm d}\Phi^+_\perp\over {\mathrm d}x}=A[x-(x_1+x_2)/2]^{c}[(x-x_1)(x_2-x)]^{d},
\label{planar7}
\ee
where $A,c$, and $d$ are constants. This model is similar to the cylindrical model, with
\be
J_\parallel={A\over R_{\rm A}}\{c(x-x_1)(x_2-x)-2d[x-(x_1+x_2)/2]^2\},
\label{planar8}
\ee
which has one sign near the center of the flux tube, at $x=(x_1+x_2)/2$, and the opposite sign nearer the edges at $x=x_1$ and $x=x_2$. The sign of $J_\parallel$ reverses at $x=x_\pm$, with
\be
x_\pm={x_1+x_2\over 2}\pm{2c\over2c+d}{x_2-x_1\over2}.
\label{planar8a}
\ee
The energy transport and the direction and return current in this model are qualitatively similar to those in the cylindrical model Equation (\ref{RC5}). In particular, both increase with increasing $d$ in Equation (\ref{planar7}), which plays a similar role to $b$ in Equation (\ref{RC5}).

The planar model facilitates incorporating many pairs of direct and return currents. For example, consider the sinusoidal potential, which satisfies the condition Equation (\ref{BC1}),
\be
{{\mathrm d}\Phi^+_\perp\over {\mathrm d}x}=b\cos\left[{x-(x_1+x_2)/2\over x_2-x_1}M\pi\right],
\label{planar9}
\ee
with $b$ a constant and $M$ an integer. This model gives
\be
J_\parallel=-{bM\pi\over R_{\rm A}(x_2-x_1)}\,\sin\left[{x-(x_1+x_2)/2\over x_2-x_1}M\pi\right].
\label{planar10}
\ee
In this case it is straightforward to integrate Equation (\ref{planar9}) to find the potential
\be
\Phi^+_\perp-\Phi^+_0={b(x_2-x_1)\over M\pi}\,\sin\left[{x-(x_1+x_2)/2\over x_2-x_1}M\pi\right]
=-{R_{\rm A}(x_2-x_1)^2J_\parallel\over M^2\pi^2},
\label{planar11}
\ee
where $\Phi^+_0$ is a constant.  For $M=1$, the direction of the FAC  for $x_1<x<(x_1+x_2)/2$ is opposite to that for $(x_1+x_2)/2<x<x_2$, and one has a single pair of direct and return FACs. The inclusion of the multiplicity factor [$M$] implies $M$ neighboring pairs of up and down FACs.

\section{Inclusion of Dissipation}
\label{s:dissipation} 

The foregoing discussion applies to undamped Alfv\'en waves propagating between the generator region and the acceleration region. Once the waves enter the acceleration region it is assumed that they are damped through some collisionless dissipation process. A self-consistency argument allows one to relate the $E_\parallel\ne0$ involved in the acceleration to the spatial decay of the waves.

\subsection{Onset of Effective Dissipation}

The onset of a flare is attributed to the turning on of effective dissipation in the acceleration region, due to  the turning on of $E_\parallel\ne0$.  The dissipation must be anomalous: this point has been recognized by many authors  \cite{AC67,C78,H85,MM87}. Although various models for the anomalous dissipation have been explored  \cite{Raadu,B93,BKB10,H12}, no consensus has emerged on the detailed microphysics involved in the creation of $E_\parallel\ne0$. An approach that avoids these details is based on the idea that the rate of anomalous dissipation is able to adjust to meet global requirements \cite{H94,M12a}. All of the proposed forms of anomalous dissipation imply highly localized and transient regions of dissipation, and effective dissipation requires a statistically large number of such localized transient regions coupled together. On a macroscopic scale, the effective rate of dissipation is determined by the statistical distribution of these localized, transient regions, and is insensitive to the details of the microphysics \cite{M12a}. In the present context, this implies that the rate of dissipation due to acceleration by $E_\parallel$ in the acceleration region can adjust to the rate of dissipation required by the rate of magnetic energy release in the energy-release region. We assume that this adjustment is relatively slow, over many Alfv\'en propagation times, in a precursor phase. We identify the onset of a flare as a threshold being reached where the coupling between the localized dissipation regions becomes effective, like a percolation threshold in a network system \cite{DM02}. In such a model, the macroscopic dissipation turns on and couples to the generator region in at most a few Alfv\'en propagation times. The power dissipated in the acceleration region can then be modeled as $R_{\rm eff}I^2$, where $R_{\rm eff}$ is an effective resistance, which adjusts to meet the required rate of dissipation.

The development of $E_\parallel\ne0$  implies that the frozen-in condition does not apply in the acceleration region, allowing the magnetic field lines to slip through the plasma. This was referred to as a ``fracture'' of the magnetic field by \citeauthor{H94} \shortcite{H94,H12}. Whereas prior to the flare, the line-tying condition precludes plasma motion in response to the stress imposed in the energy-release region, once $E_\parallel\ne0$ develops in the acceleration region, the plasma above this region can move relative to the plasma below it. This motion corresponds to the fluid motion [${\bi u}^\pm$] in the Alfv\'en waves, and is essential in allowing Alfv\'enic transport of energy. The assumption that the potential surfaces all close within the acceleration region is an oversimplification, that ignores the requirement that a return current be set up between the acceleration region and the photosphere, or more specifically, the region of cross-field current closure. As in the application to auroral acceleration \cite{Ma09}, this requires some S-shaped, as well as U-shaped potentials. Once effective dissipation turns on, the rates of energy release and dissipation adjust to balance each other within a few Alfv\'en propagation times.

\subsection{Impedance Matching}

The foregoing arguments suggest that the global requirement on the dissipation and electron acceleration is that Alfv\'enic flux released in the energy-generation region be completely absorbed in the acceleration region. This condition corresponds to impedance matching.

Suppose the dissipation is described by an effective (anomalous) conductivity, $\sigma_{\rm eff}$, which is averaged over the distribution of localized dissipation regions. Let $1/R_{\rm eff}$ be the effective conductivity  integrated over height in the acceleration region. The condition for zero reflection of an Alfv\'en wave incident on the acceleration follows by analogy with Equation (\ref{AW9}):
\be
r_{\rm ref}={R_{\rm eff}-R_{\rm A}\over R_{\rm eff}+R_{\rm A}}=0.
\label{perfect1}
\ee
It follows that all of the incoming energy is dissipated completely when the effective resistance of the acceleration region adjusts such that it is equal to the Alfv\'enic impedance. This impedance matching condition [$R_{\rm eff}=R_{\rm A}$] leads to a model for the dissipation that is independent of the details of the microphysics involved in the anomalous dissipation. 

\subsection{ Dissipation and the Parallel Electric Field}

Outside the acceleration region, one can assume $E_\parallel^\pm=0$ in the waves, so that the field lines are equipotentials, and the frozen-in condition applies. Within the acceleration region $E_\parallel^+\ne0$ implies that neither of these conditions apply. The plasma slips through the magnetic field, such that all fields associated with the wave, $E_\perp^+$, $\Phi_\perp^+$, \etc,   decrease with decreasing $z$. In particular, this implies that $\Phi_\perp^+$ has a dependence on $z$ that is not included in the dependence through $z\pm v_0t$. This leads to a component of the potential electric field along $z$:
\be
E^+_\parallel=-\left({\partial\over\partial z}\mp{1\over v_0}{\partial\over\partial t}\right)
\Phi_\perp^+.
\label{AR1}
\ee
This electric field is assumed to accelerate particles, transferring energy from the waves to the particles, leading to the damping of the waves. 

\begin{figure} [t]
\centerline{
\includegraphics[scale=0.5]{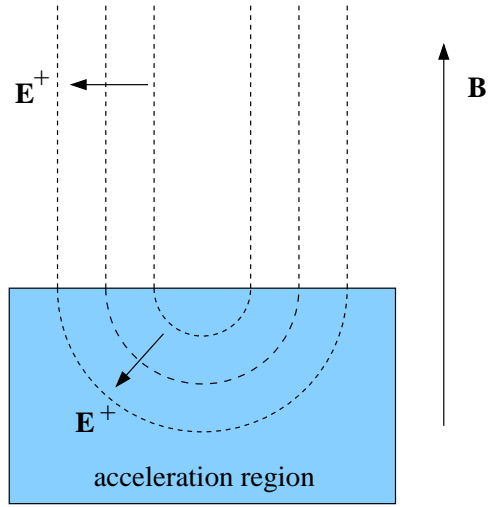}
}
\caption{The lines of constant potential are along vertical field lines above the acceleration region, and are shown to close as semicircles within the acceleration region. The potential gradient changes from horizontal to vertical with decreasing height.
}
\label{fig:pot}
\end{figure}

The conventional interpretation of how $E_\parallel\ne0$ is set up was suggested by Gurnett in 1972 \cite{A77}, and is illustrated here in Figure~\ref{fig:pot}. The idea is that the equipotential surfaces, which are along field lines above the acceleration region, must close across field lines within the acceleration region. Then $E^+_\parallel$ is attributed to the normal to equipotential surfaces having a component along field lines. A physical argument is that the equipotential surfaces must close above a perfectly conducting boundary, and that the fluid velocity associated with the waves must go to zero above the line-tied boundary; that is, both ${\bi E}_\perp$ and ${\bi u}_\perp$ must go to zero above the photosphere. The following analytic model illustrates how this leads to $E_\parallel\ne0$.

In a cylindrical model, above the acceleration region the field lines are equipotentials, so that $\Phi^\pm$ depends only on $r$ and $z\pm v_0t$. Within the acceleration region, it is assumed that the equipotential surfaces close across the field lines. Suppose the acceleration region is at $z<z_t$. The shape of an equipotential surface at $z<z_t$ can be described as a function of $r,z$. Let the potential surfaces be described by the equation $\Phi(r,z)-\Phi_0=f(r,z)=0$, where different surfaces depend (implicitly) on a constant $r_t$, that is  the value of $r$ at $z\ge z_t$. The ratio of the parallel and perpendicular components is then determined by the function $f(r,z)$:
\be
E_\parallel^+:E_\perp^+={\partial f(r,z)\over\partial z}:{\partial f(r,z)\over\partial r}.
\label{AR2}
\ee
The lower boundary of the acceleration region is the surface $f(r,z)=0$ for $r_t=R$, where $r_t=R$ is the boundary of the cylinder at $z=z_t$. A particle that passes through the acceleration region along a given field line at radius $r$ is accelerated through a potential difference
\be
\Delta\Phi^+=|\Phi^+(R)-\Phi^+(r)|,
\label{AR3}
\ee 
where the right hand side is to be evaluated at $z\ge z_t$.

A specific model for the surfaces is to assume that they are elliptical. One has vertical equipotential surfaces [$r=r_t<R$, for $z>z_t$] and elliptical surfaces,
\be
f(r,z)=r^2+\alpha^2(z_t-z)^2-r_t^2=0, 
\label{AR4}
\ee
for $z<z_t$, with $\alpha$ the ratio of the semi-axes of the ellipse. The case $\alpha=1$ corresponds to semi-circular surfaces illustrated in Figure~\ref{fig:pot}. The lower boundary of the acceleration region is the surface $r^2+\alpha^2(z_t-z)^2-R^2=0$. The ratio Equation (\ref{AR2}) becomes
\be
E_\parallel^+:E_\perp^+=\alpha^2(z-z_t):r.
\label{AR5}
\ee
The boundary condition Equation (\ref{BC1}) implies that both $E_\parallel^+$ and $E_\perp^+$ vanish on (and below) the lower boundary of the acceleration region.

While plausible, this model is heuristic. The assumptions made is writing down the wave solution Equation (\ref{AW1}) and introducing the wave potential Equation (\ref{AW1a}) are not valid for $E_\parallel\ne0$. 

\section{Discussion}
\label{s:discussion} 

The three features of the Alfv\'en wave model emphasized here are energy transport, the mapping of potential differences, and the setting up of intrinsically new current loops involving FACs. In this section, the cylindrical and planar models described in Section~\ref{s:cylindrical}  are used to illustrate these features.

\subsection{Alfv\'enic Energy Transport}

Energy transport by Alfv\'en waves is well understood: energy propagates at $v_0\approx v_{\rm A}$ along the direction of the background magnetic field. Such energy propagation has been invoked in several different flare models. For example, \inlinecite{P74} developed a model in which the Alfv\'en waves transport the energy into the corona from below the photosphere during the flare. In the flare model due to \inlinecite{FH08}, and in the model developed here, energy already stored in the corona is converted into an Aflv\'enic energy flux in a generator region, is transported downwards, and is transferred to energetic electrons in an acceleration region. 

In the Alfv\'en wave model, the energy transport is closely linked to the FAC [$J_\parallel$] in the waves. It is of interest to compare the present model for energy transport with an earlier model \cite{M12b} in which, prior to the flare, the FAC profile in a cylindrical flux tube is described by a function $j_1(\xi)$, with $\xi=r/R$. This profile changes, to $j_2(\xi)$, after passage of an Alfv\'enic front launched by the onset of the flare. It was found that power transported increases with increasing concentration of $j_2(\xi)$ towards $\xi=0$; this increase continues as the central concentration of $j_2(\xi)$ increases to arbitrarily large values, compensated by a return current flowing at larger radii. In the present model, the pre-flare FAC is excluded by the assumption that the field lines are straight. This may be interpreted as assuming that the guiding magnetic field is arbitrarily strong, and that both the pre-flare FAC and the FAC in the Alfv\'en waves may be treated as (independent) perturbations. The Alfv\'en wave model then has the same qualitative and semi-quantitative features as in the earlier model. Specifically, in the cylindrical case, the power transported by the Alfv\'en waves increases as $J_\parallel^\pm$ becomes increasingly concentrated towards the axis of the cylinder, where it enhances the pre-flare FAC, with $J_\parallel^\pm$ in the opposite direction to the pre-flare FAC at larger radii. In particular, for the model Equation (\ref{RC5}) both the concentration of the FAC and the energy flux increase with increasing power-law index $b$. The planar model Equation (\ref{planar8}) has similar properties, with the power-law index $d$ the counterpart of $b$ in the cylindrical model. 

\subsection{Transport of Potential}

A new feature of the Alfv\'en wave model in the solar context is the transport of potential. The Alfv\'en waves are launched in the generator region, where the power released is due to a cross-field potential and a cross-field current: the cross-field potential in the wave matches that at the boundary of the generator region, and the FAC in the wave arises from redirection of the cross-field current at the boundary of the generator region. Which of these two boundary conditions dominates determines whether the generator can be approximated as constant-voltage or constant-current, which in turn depends on the ratio of the internal and external impedances. The Alfv\'en waves transport the potential along field lines to the acceleration region. Reflection of the Alfv\'en waves from both regions modifies the potential, providing a feedback that regulates the rate of dissipation in the acceleration region and the rate of energy release in the generator region.

A realistic theory requires a physical model to determine the form of the wave potential. Here, simple analytical models are chosen to satisfy the boundary condition Equation (\ref{BC1}) at the edge of the flux tube within which the Alfv\'en waves are confined. The magnitude of the potential is not constrained by the theory, but it is implausible that the total potential available [$\Phi_{\rm tot}=10^9\,-\,10^{10}\,$V] can be transported by a single Alfv\'en wave. There is no evidence that $\Phi_{\rm tot}$ appears across the acceleration region; the evidence is that the potential that does appear across the acceleration region is $\Delta\Phi=\Phi_{\rm tot}/M$, with $M$ of order $10^6$. 

\begin{figure} [t]
\centerline{
\includegraphics[scale=0.4]{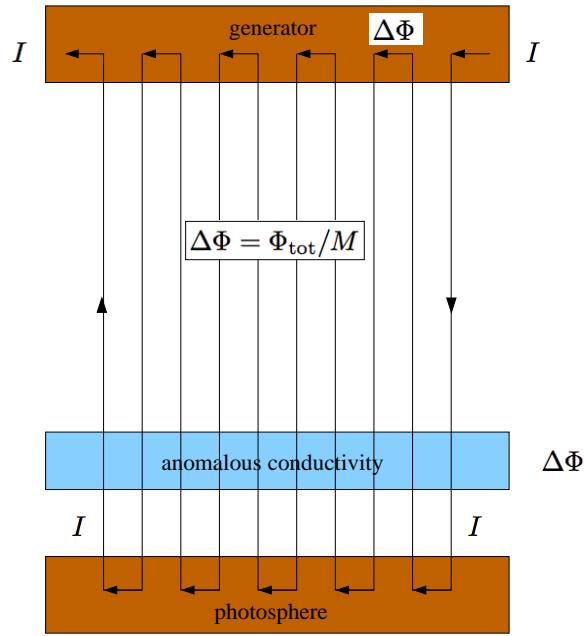}
}
\caption{The model envisaged here has the generator region at the top, with a total potential $\Phi$ and cross-field current $I$, below which is an energy transport region to an acceleration region with anomalous conductivity. The cross-field current is redirected along field lines to close in a conducting boundary, which is assumed to be the photosphere. Following Holman (1985), the current [$I$] is assumed to flow up and down multiple [$M$] times, forming a picket-fence structure. The potential across the acceleration region is $\Delta\Phi=\Phi/M$.
}
\label{fig:picket}
\end{figure} 

The explanation for the multiplicity [$M$] proposed by \inlinecite{H85} can be incorporated into the model. \citeauthor{H85}'s suggestion is that there are $M$ pairs of up and down current channels, leading to the ``picket-fence'' model illustrated in Figure~\ref{fig:picket}. One requires that the potential across the acceleration region [$\Delta\Phi=\Phi_{\rm tot}/M$] be of order $10^4\,$V  for each channel, each of which has up and down currents [$\Delta I$] of order $I$. It is unrealistic to use cylindrical geometry to describe multiple pairs of up and down currents, which would occur in alternate concentric rings. In a planar geometry, the pairs of up and down current are in alternate current sheets. A more realistic geometry requires an arrangement for the $M$ current pairs, in either cylindrical ($r,\phi$) or cartesian ($x,y$) geometry, combined with a model for each of the current pairs. 

A physical explanation is needed for the multiplicity [$M$] of current pairs. One possibility is to consider restrictions imposed by the microphysics on the potential in the Alfv\'en waves. If the potential is restricted to $\Phi^+\lesssim10^4\,$V, then $M=10^6$ current channels must develop in order to provide the required total rate of energy loss. Moreover, these current channels must develop simultaneously; if they develop sequentially, the implied timescale of $10^6$ Alfv\'en propagation times greatly exceeds the timescale of a flare. These problems are not discussed further here.

\subsection{Dissipation and Acceleration}

How Alfv\'en waves damp and accelerate particles effectively is an unsolved problem. Some form of anomalous dissipation is required. As has long been recognized, a requirement for a current-driven instability is a very high current density, which implies filamentation of the current into many high-current-density channels  \cite{H85,T85,MM87,M90,EH95,T95}. Let the current threshold for instability be $J_{\rm crit}$. Suppose each filament is cylindrical, and of radius $\lambda$, such that the current in each filament is $I_{\rm f}=\pi\lambda^2J_{\rm crit}$. One requires a large number [$N_{\rm f}=I/I_{\rm f}$] of such channels. For a specific model, \inlinecite{H12} estimated $J_{\rm crit}=4\times10^3\rm\,A\,m^{-2}$ and $\lambda=1\,$meter. An unavoidable implication is that anomalous dissipation requires that the dissipation region is highly structured, involving very small scales perpendicular to the field lines. The following argument based on the Alfv\'en wave model leads to a similar conclusion.

Suppose one balances the incoming Poynting vector in the Alfv\'en waves with the outgoing kinetic energy flux in accelerated electrons with a number density $n_e$. This gives
\be 
{1\over\mu_0v_0}\left({d\Phi^+(r)\over dr}\right)^2=n_e\left({2\over m}\right)^{1/2}
[e\Delta\Phi^+(r)]^{3/2},
\label{sc1}
\ee
with $\Delta\Phi^+(r)$ given by Equation (\ref{AR3}), and where $[2e\Delta\Phi^+(r)/m]^{1/2}$ is the speed of the accelerated electrons. Using Equation (\ref{AR3}), one may integrate Equation (\ref{sc1}) to find
\be 
e\Delta\Phi^+(r)=2mv_0^2\left({r\over\lambda}\right)^4,
\qquad {1\over\lambda}={\omega_p\over4c},
\label{sc2}
\ee
and use Equation (\ref{AW5}) to find
\be 
|J_\parallel^+|=en_ev_0{3\over2}\left({r\over\lambda}\right)^2.
\label{sc3}
\ee
With $e\Delta\Phi^+(r)$ of order $10^3\times mv_0^2$ under coronal conditions, it follows from Equation (\ref{sc2}) that the model requires perpendicular structure in $\Phi^+(r)$ on a scale of a few tens of skin depths [$c/\omega_p$] which is of the same order of magnitude as the $\lambda$ estimated by \inlinecite{H12} by a different argument. It follows from Equation (\ref{sc3}) that there is a high current density, sufficient to trigger anomalous conductivity, on the scale $\lambda$.

The result, Equation (\ref{sc3}), demonstrates an inconsistency in the Alfv\'en model, as developed here. The Alfv\'en wave model applies on a macro scale, plausibly describing the energy and potential transport to the top of the acceleration region. Effective dissipation involves micro-scale structures, and these must be taken into account within the dissipation region. The model developed here for the dissipation region can at best be regarded as applying to properties averaged over a statistically large distribution of such micro-scale structures. 

\subsection{Return Current}

The long-standing ``number problem'' can be partly resolved by requiring a return current between denser regions of the solar atmosphere and the acceleration region. Existing models for the return current have invoked both electrostatic and inductive effects \cite{BB84,SS84,vdO90}. The Alfv\'en wave model provides a new way of modeling the return current, in terms of Alfv\'en waves setting up new current loops, that close across field lines in the acceleration region and in denser regions of the solar atmosphere. Such a model has an important qualitative difference from these earlier models, in which the ions were neglected \cite{vdO90}. Neglecting the ions leads to neglecting the polarization current, so that the inductive effects propagate effectively at the speed of light. It is essential to include the polarization current, and then Equation (\ref{pd3}) implies that inductive effects are transported along field lines by Alfv\'en waves at the MHD speed. An Alfv\'en wave model for an inductively driven return current is needed to show how the electrons are continuously resupplied to the acceleration region, as is necessary for both ``first phase'' electron acceleration and for auroral electron acceleration \cite{Ma09}.

\section{Conclusion}
\label{s:conclusion} 

The main point of this article is that large-amplitude Alfv\'en waves play an important role in solar flare physics. These are not ``waves'' in the conventional sense, with a well-defined frequency and wave vector, but are specific solutions of the Alfv\'en wave equation Equation (\ref{MEperpt}) involving a cross-field wave potential [$\Phi_\perp^\pm$] and field aligned currents, $J_\parallel^\pm$, determined by the wave potential. Energy transport is along field lines at the MHD speed [$v_0\approx v_{\rm A}$] as for any (torsional) Alfv\'en wave. The wave potential provides the mapping of the potential imposed in the generator region onto the acceleration region. Reflected Alfv\'en waves transport the reaction (of the acceleration region or the photosphere) back to the generator region, providing a coupling between them. The Alfv\'en waves set up new closed current loops, with oppositely directed FACs closing across field lines at the two ends (in the generator region and the photosphere). 

A long-standing problem with models involving particle acceleration by Alfv\'en waves is the mechanism that allows the Alfv\'en waves to damp and to transfer their energy to particles. The interpretation of the wave amplitude in terms of a cross-field potential provides a way of interpreting this damping without identifying the microphysics involved. Within the acceleration region it is assumed that the wave amplitude, and hence the cross-field potential, decreases downwards. There is then a nonzero gradient of the potential along the field lines implying a nonzero $E_\parallel$. This $E_\parallel$ accelerates the particles. In an idealized model, the Alfv\'enic energy flux at the top of the acceleration region is converted into a kinetic-energy flux in accelerated electrons at the bottom of the acceleration region, where the wave amplitude (potential) is zero. The rate of dissipation can be determined by assuming that the anomalous dissipation processes adjust to maximize the power dissipated. This leads to the impedance matching condition for the effective resistance, $R_{\rm eff}=R_{\rm A}$, of the acceleration region. All anomalous dissipation processes involve microphysics on tiny space scales, of order one meter  in the present context \cite{H12}, and the relation between processes on micro and macro scales is treated heuristically here. This relation needs to be explained in a more detailed model. 

The Alfv\'en wave model indicates how long-standing, unsolved problems related to bulk energization of electrons may be resolved. It is assumed that the acceleration results from $E_\parallel\ne0$ in an acceleration region, along field lines above a hard X-ray source in the chromosphere. One problem is the seeming inconsistency between the large cross-field potential, of order $10^{10}\,$Volts, and the energy, $e\Delta\Phi$ of order $10^{4}\,$eV, of the accelerated electrons. One way of resolving this \cite{H85} is to assume that the pre-flare current flows back and forth between the generator and dissipation regions many ($M=10^6$) times. Electrons are accelerated downward in a sequence of $M$ jumps in potential, one jump for each instance of $I$ flows up through the acceleration region. The additional insight that the Alfv\'en wave model provides is that $M$ current paths may be regarded as $M$ closed current loops set up by Alfv\'en waves. The number problem is resolved by identifying the total rate [$\,{\dot{\!N}}$] that electrons precipitate as $M$ times the rate [$I/e$] for each of these loops. However, there is no obvious explanation within the model for the value of the multiplicity [$M$]. A speculation is that the microphysics in the acceleration region constrains the value of $e\Delta\Phi$. 

There are unsolved problems related to the acceleration region \cite{H12}. In particular, the self-consistency argument for $E_\parallel\ne0$, based on the damping of the waves due to energy transfer to particles by $E_\parallel$, is heuristic, and needs to be complemented by a specific mechanism that accounts for $E_\parallel\ne0$. All such mechanisms appear to involve very small scales perpendicular to the field lines, estimated to be of order one meter by \inlinecite{H12}. An outstanding problem is how this microphysics relates to the macroscopic model discussed in this article. 

The Alfv\'en wave model developed here is highly simplified, and is applied specifically only to transport of energy between the generator and acceleration region. However, the physics involved is likely to be widely relevant to energy transport and energy release in the solar atmosphere.

%

%

%
\appendix   

\section{Response of a Plasma at Low Frequencies}
\label{A-appendix}

The response of a cold plasma may be described by the dielectric tensor $K_{ij}(\omega)$ \cite{Stix}:
\be
D_i(\omega)=\varepsilon_0K_{ij}(\omega)E_j(\omega),
\qquad
K_{ij}(\omega)=
\left(\begin{array}{ccc}
S&-iD&0\\
iD&S&0\\
0&0&P
\end{array}
\right).
\label{pd3a}
\ee
At sufficiently low frequencies, when dissipation is neglected, one has
\be
S\approx1+{c^2\over v_{\rm A}^2},
\qquad
D\approx0,
\qquad
P=1-{\omega_p^2\over\omega^2}.
\label{pd4}
\ee
The electric induction [${\bi D}$] includes the response of the plasma through the polarization [${\bi P}$]:
\be
{\bi D}=\varepsilon_0{\bi E}+{\bi P},
\qquad
\rho_{\rm ind}=-\div{\bi P},
\qquad
{\bi J}_{\rm ind}={\partial{\bi P}\over\partial t},
\label{pd5}
\ee
where $\rho_{\rm ind}$ and ${\bi J}_{\rm ind}$ are the induced charge and current densities. The perpendicular component of the response is
\be
{\bi P}_\perp={c^2\over v_{\rm A}^2}\varepsilon_0{\bi E}_\perp.
\label{pd6}
\ee
The temporal derivative of Equation (\ref{pd6}) gives
\be
{\bi J}_{\rm ind \perp}={c^2\over v_{\rm A}^2}\varepsilon_0{\partial{\bi E}_\perp\over\partial t},
\label{pd7}
\ee
which reproduces Equation (\ref{pd3}). The parallel term in the low-frequency, cold-plasma limit gives
\be
{\partial J_{\rm ind\parallel}\over\partial t}=\varepsilon_0\omega_p^2E_\parallel.
\label{pd8}
\ee

The components Equation (\ref{pd7}) and  Equation (\ref{pd8}) of the current density are included in Maxwell's equations. For the perpendicular components, one obtains
\be
\left({1\over v_0^2}{\partial^2\over\partial t^2}-\nabla^2
\right){\bi E}_\perp
=-{{\bf\nabla}_\perp\rho_{\rm ext}\over\varepsilon_0}
-\mu_0{\partial{\bi J}_{\rm ext\perp}\over\partial t},
\label{MEperp}
\ee
where the right hand side includes source terms. The result Equation (\ref{MEperp})
reproduces Equation (\ref{MEperpt}) when the source terms are neglected. For the parallel component, one obtains
\be
\left[{1\over c^2}{\partial^2\over\partial t^2}+{\omega_p^2\over c^2}-\nabla^2
\right]E_\parallel=-{1\over\varepsilon_0}{\partial\rho_{\rm ext}\over\partial z}
-\mu_0{\partial J_{\rm ext\parallel}\over\partial t}.
\label{MEpar}
\ee

%
%

%
%
%

\end{article} 
\end{document}